# Ultrawide Dynamic Bandwidth Modulation of an Antiresonant Nanoweb Hollow-Core Fiber

Ricardo E. da Silva and Cristiano M. B. Cordeiro

***Abstract*— We experimentally demonstrate an acoustically modulated antiresonant nanoweb hollow-core fiber (N-HCF) for the first time. The N-HCF contains two off-center air cores with a diameter difference of 5 µm, separated by a nanoweb of silica. We analytically simulate the influence of the N-HCF's core diameter, cladding wall, and nanoweb thicknesses on the confinement losses, effective indices, and beatlengths of the guided fundamental ($HE_{11}$) and higher-order modes ($TE_{01}$, $TM_{01}$), from 750 to 1200 nm. The phase-matching of the acoustic waves and modal beatlengths is also estimated and discussed. The fabricated 3.6 cm long acousto-optic device modulates record-wide bandwidths (up to 450 nm) while providing high modulation depths (up to 8 dB) at low drive voltages (10 V). Simulated and measured results provide useful insights for designing, modeling, and characterizing the N-HCF's transmission spectrum and modulation performance. These achievements lead to highly efficient, compact, and fast all-fiber sensors and modulators promising for application in pulsed fiber lasers.**

***Index Terms*— Acousto-optic devices, antiresonant nanoweb hollow-core fiber, analytical modeling, flexural acoustic waves.**

## I. Introduction

ALL-FIBER[1] acousto-optic modulators (AOMs) enable outstanding electrically-tunable spectral filters, fiber sensors, mode-locked and Q-switched pulsed fiber lasers [1], [2], [3]. Thus, AOMs contribute to reducing losses, the number of components, device size, and cost when compared to the traditional free-space bulk modulators, providing short switching times, integration with other fiber components, and dynamic tuning of spectral and power properties.

An AOM can be fabricated by combining a piezoelectric transducer (PZT), an acoustic horn, and an optical fiber segment. The AOM generates flexural acoustic waves which couple the power between the guided optical modes in the fiber, modulating a notch band in the fiber transmission spectrum. The notch modulation depth and center wavelength are usually tuned respectively by the drive electrical voltage (power) and frequency. However, the weak overlap of the generated acoustic wave and the optical modes in the core of standard solid fibers considerably decreases the acousto-optic efficiency. This decreased efficiency usually requires etched, tapered, or long fibers, or even high voltages to increase the modulation depth [2], [4], [5], [6], [7]. Similarly, the modulated notch bandwidth is usually a few nanometers in standard fibers, limiting the modulated bandwidth and maximum peak power of fiber laser applications. As a solution, AOMs using short fibers have been employed to increase the modulated width at the expense of decreased modulation depth and efficiency. Overall, broad modulated widths (30 – 362 nm) have still been achieved by employing AOMs with reduced fiber diameters (80 – 21 µm) [6], [8], [9].

Here, we experimentally demonstrate the highly efficient interaction of a dual-core nanoweb HCF (N-HCF) and flexural acoustic waves. We show that the N-HCF provides the widest modulated bandwidth when compared to other fibers used in previous studies. The N-HCF is analytically simulated, and the properties of the guided modes in the fiber cores are investigated in detail (Section II). The effect of the N-HCF's geometry and dimensions on acousto-optic performance is analyzed. Fabrication of a compact AOM is described, and the N-HCF modulated spectrum is characterized from 750 to 1200 nm (Section III). The results show high modulation efficiencies using low drive voltages and short fiber length (Section IV).

## II. Analytical Modeling and Study of the Acousto-Optic Properties of the N-HCF

Fig. 1(a) shows the N-HCF cross-section indicating details of the fiber cores and nanoweb in Fig. 1(b). Fig. 1(c) illustrates the modeled fiber, materials, and dimensions (outer and inner diameters, $D_e$, $D_i$, silica wall thickness, $d$, and the nanoweb thickness, $t$). We have characterized the N-HCF's dimensions by using a scanning electron microscope (SEM). The fiber parameters are derived directly from high-resolution images, similar to those shown in Fig. 1(a) and 1(b). The inset in Fig. 1(b) shows a nanoweb's zoomed region used to provide better accuracy for measuring the nanoweb thickness $t$.

The N-HCF is modeled based on the antiresonant reflecting optical waveguide (ARROW) formulation, which has been extensively studied to simulate planar (slab) and fiber waveguides with distinct core geometries [10], [11], [12], [13], [14]. Generally, these analytical models consider the guided modes mostly confined to the inscribed circle formed by the silica edges of the air core (with a radius, $r = D/2$). This is useful

[1]This work was supported by the grants 2022/10584-9, 2024/02995-4, São Paulo Research Foundation (FAPESP), 309989/2021-3, Conselho Nacional de Desenvolvimento Científico e Tecnológico (CNPq).

R. E. da Silva (resilva@unicamp.br) and C. M. B. Cordeiro (cmbc@ifi.unicamp.br) are with the Institute of Physics Gleb Wataghin, University of Campinas (UNICAMP), Campinas, 13083-859, Brazil.

for estimating the modes' effective indices and confinement loss in circular, triangular, squared, and polygonal-shaped fiber cores [10], [12]. The confinement losses increase with the number of core silica edges, while the core shape negligibly affects the wall resonances in the transmission spectrum. In contrast, the resonance wavelengths, $\lambda_l$, strongly depend on the wall thickness $d$, given as [10],

$$\lambda_l = \frac{2d}{l}\sqrt{{n_s}^2 - {n_a}^2} \quad (1)$$

where, $l$, is the resonance order, $n_s$ and $n_a$, are respectively the refractive indices of the silica and air regions ($n_a = 1$, and $n_s$ is calculated using Sellmeier formulation [6]).

We have modeled the N-HCF considering that the guided modes are mostly confined in the inscribed circles in the cores (dashed red and blue circles in Fig. 1(c)), by using the method described in [10]. Fig. 1(d) and 1(e) illustrate the modeled cores with different core diameters, $D_{C1}$ and $D_{C2}$ (core 2 is 5 µm larger than core 1). Two studies are set to investigate separately the contribution of wall and nanoweb on the modal guidance: the first considers the mode fields overlapping only with the wall ($d_1 = 36$ µm), while in the second, the mode overlaps only with the nanoweb ($d_2 = 712$ nm).

The real $n_{eff\,(\mathrm{RE})}$ and imaginary $n_{eff\,(\mathrm{IMG})}$ parts of the effective indices of the fundamental, $HE_{11}$, and higher-order modes, $TE_{01}$ and $TM_{01}$, are estimated considering the following analytical solutions [14],

$$n_{eff(\mathrm{RE})} = n_a\left[1 - \frac{j^2}{2k_a^2r^2} - \frac{j^3}{k_a^3r^3}\frac{\cot\phi}{\sqrt{\epsilon - 1}}\right]\cdot\begin{cases} 1 & \mathrm{TE}_{0n} \\ \epsilon & \mathrm{TM}_{0n} \\ \dfrac{(\epsilon + 1)}{2} & \mathrm{HE}_{m,n} \end{cases} \quad (2)$$

$$n_{eff(\mathrm{IMG})} = n_a\frac{1 + \cot^2\phi}{\epsilon - 1}\frac{j^3}{k_a^4r^4}\cdot\begin{cases} 1 & \mathrm{TE}_{0n} \\ \epsilon^2 & \mathrm{TM}_{0n} \\ \dfrac{(\epsilon^2 + 1)}{2} & \mathrm{HE}_{m,n} \end{cases} \quad (3)$$

where,

$$j = \begin{cases} j_{1n} & \mathrm{TE/TM}_{0n} \\ j_{m-1,n} & \mathrm{HE}_{m,n} \end{cases} \quad (4)$$

are the zeros of the Bessel function ($j_{01} = 2.405$ for $HE_{11}$, $j_{11} = 3.832$ for $TE_{01}$ and $TM_{01}$ [11]), $m$ and $n$, are the mode azimuthal and radial indices, $\epsilon = n_s^2/n_a^2$ is the relative permittivity, $k_a = 2\pi n_a/\lambda$ is the wavenumber, and $\phi = k_a d(n_s^2 - n_a^2)^{1/2}$, is the phase change caused by the mode propagation from the inner to the outer surface of the silica wall.

We have simulated the effective index, $n_{eff\,(\mathrm{RE})}$, and the confinement loss coefficient, $\alpha = 8.69 k_a n_{eff\,(\mathrm{IMG})}$ [10] of the modes $HE_{11}$, $TE_{01}$, and $TM_{01}$, from $\lambda = 750$ to 1200 nm, using respectively (2) and (3). Fig. 2(a) and 2(b) show the losses in the N-HCF cores with the silica wall and nanoweb. The inset in Fig. 2(a) shows a detail of the wall resonances for $d_1 = 36$ µm (for simplicity, only the minimum losses indicated in the inset are shown in Fig. 2(a) and 2(b)). We note that the losses in core 2 are lower than those in core 1 (about 2.5 times lower for $HE_{11}$ at 980 nm). In general, $HE_{11}$ shows reduced losses compared to $TE_{01}$ and $TM_{01}$. The nanoweb induces a broad attenuation resonance at shorter wavelengths approaching 750 nm. In contrast, wall and nanoweb losses overlap near the spectrum center (~ 980 nm), indicating negligible effect of different silica thicknesses around the cores. Similarly, $d$ has a low influence on $n_{eff\,(\mathrm{RE})}$, as shown in Fig. 2(c) and 2(d) (wall and nanoweb curves agree well deviating at resonances). The optical beat length of the fundamental $HE_{11}$ and higher-order modes, $TE_{01}/TM_{01}$, at the wavelength $\lambda_C$, is given as [4],

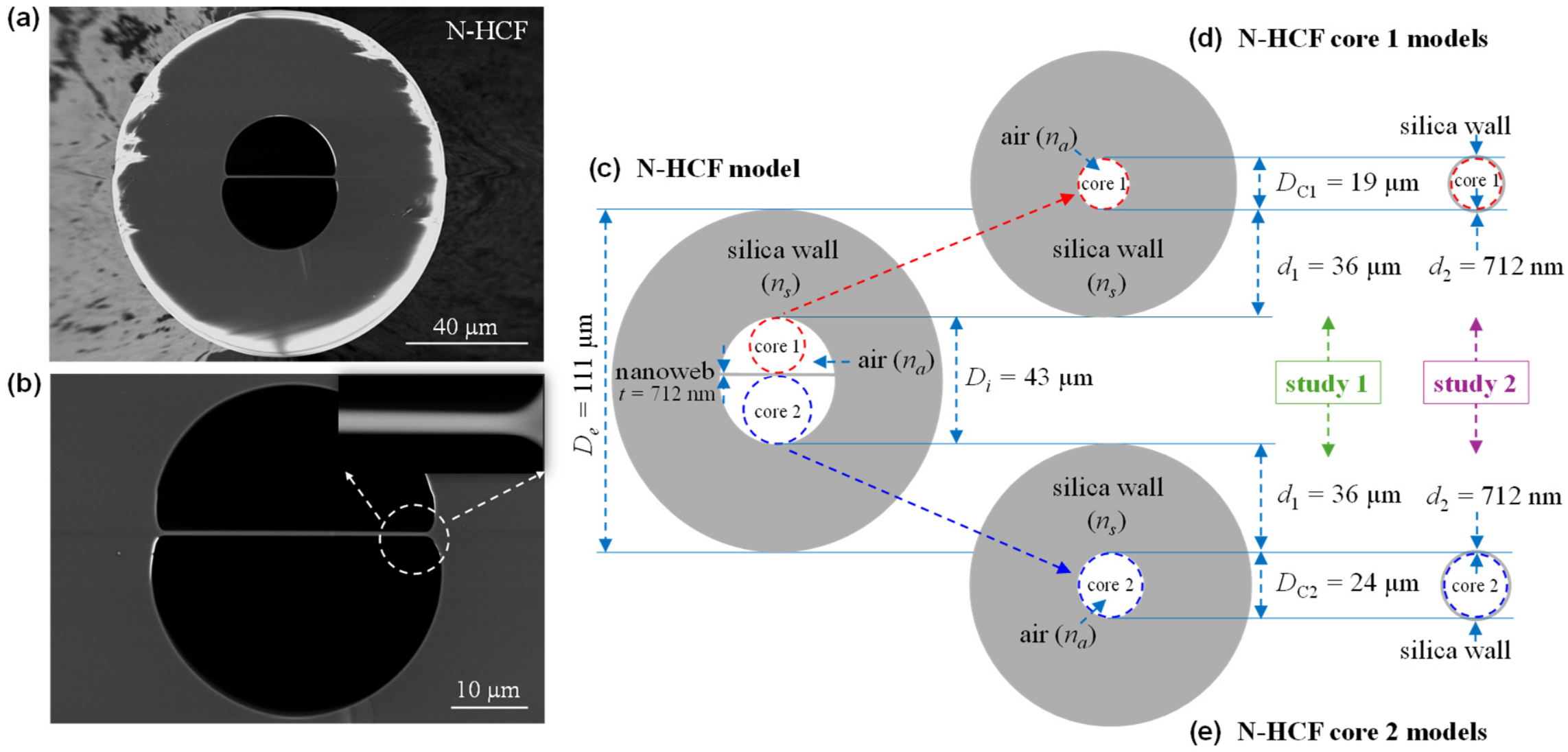

**Fig. 1**. (a) Cross-section of the antiresonant nanoweb hollow-core fiber (N-HCF) with a (b) detail of the air cores and nanoweb (inset). (c) Illustration of the N-HCF's model, materials, and dimensions used in the analytical simulations (outer $D_e$ and inner $D_i$ diameters, silica wall thickness $d$, and nanoweb thickness $t$). The guided modes are mostly confined to the inscribed circles in the air cores (dashed red and blue circles). (d) Core 1 and (e) core 2 are modeled separately to investigate the influence of the core diameters, $D_{C1}$, $D_{C2}$, silica wall, and nanoweb thickness ($d = t$) on theo modal properties. $n_s$ and $n_a$ are the silica and air refractive indices.

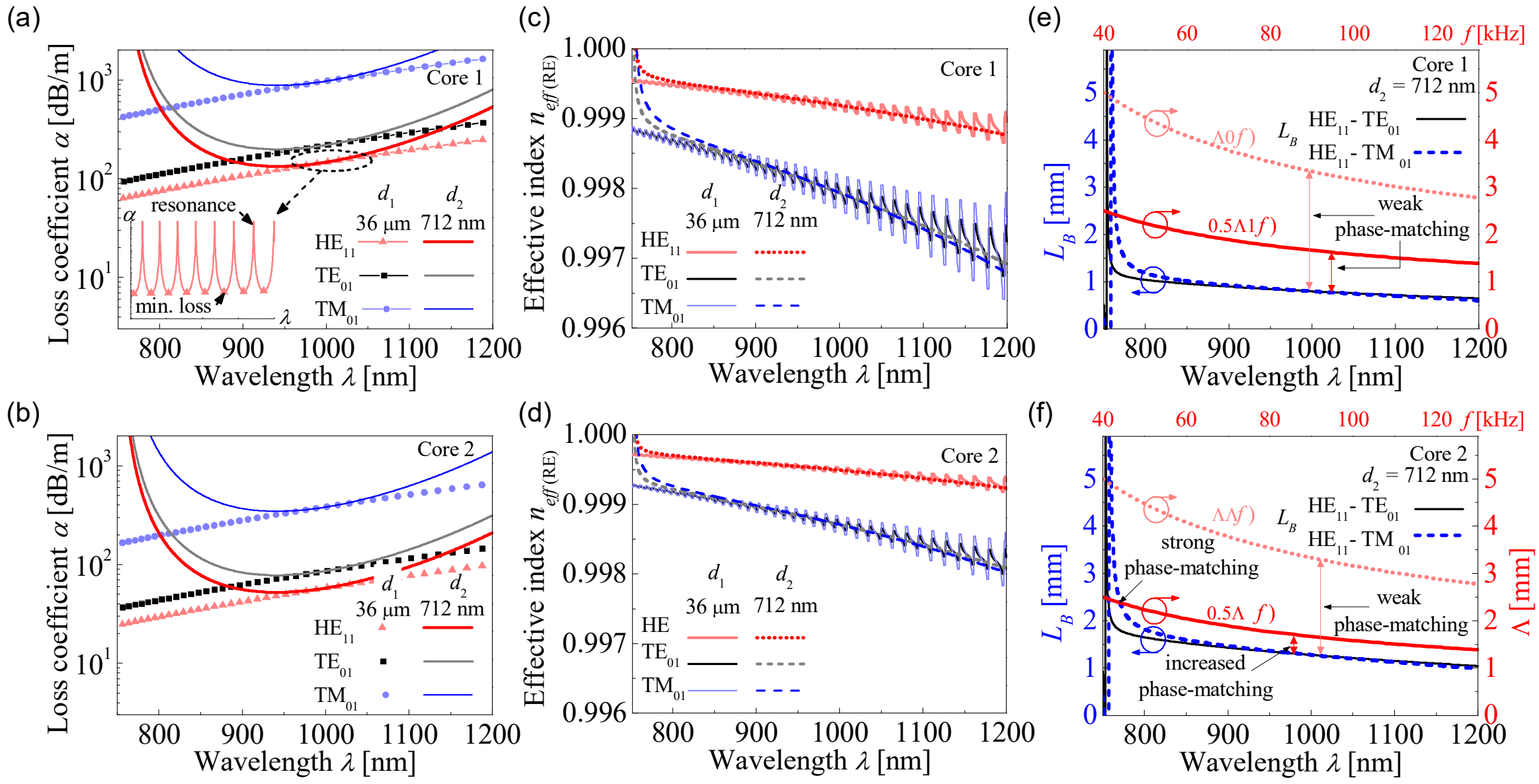

**Fig. 2**. Simulation of the (a)(b) confinement loss coefficient $\alpha$, (c)(d) real part of the effective refractive index $n_{eff\,(\mathrm{RE})}$, and (e)(f) beatlength $L_B$ of the fundamental mode HE$_{11}$ and higher-order modes TE$_{01}$ and TM$_{01}$ in the cores of the N-HCF ($L_B$ is compared to the acoustic period $\Lambda$ and its first harmonic 0.5$\Lambda$). The parameters are estimated considering the cores surrounded by the silica wall ($d_1$= 36 μm thickness) or nanoweb ($d_2$ = 712 nm thickness). The inset in (a) shows a detail of the wall resonances for $d_1$ = 36 μm.

$$L_B = \frac{\lambda_C}{n_{\mathrm{HE}} - n_{\mathrm{TE/TM}}}, \tag{5}$$

where, $n_{\mathrm{HE}}$ and $n_{\mathrm{TE/TM}}$, are the $n_{eff\,(\mathrm{RE})}$ effective indices of HE$_{11}$, TE$_{01}$ (or TM$_{01}$), as shown in Fig. 2(c)-(d).

Flexural acoustic waves change the optical path length of the guided modes in the air core, efficiently coupling power between the fundamental and higher-order modes when their beatlength matches in phase the acoustic period ($L_{\mathrm{B}} = \Lambda$), given as [15],

$$\Lambda = \left(\frac{\pi D_e c_{ext}}{2f}\right)^{\frac{1}{2}}, \tag{6}$$

where, $D_e$ = 111 μm, is the fiber outer diameter and, $c_{ext}$ = 5740 m/s, is the extensional acoustic velocity [16]. Thus, the resonant $\lambda_{\mathrm{C}}$ can be tuned by the acoustic frequency $f$ in (6) as,

$$\lambda_{\mathrm{C}} = \left(n_{\mathrm{HE}} - n_{\mathrm{TE/TM}}\right)\Lambda. \tag{7}$$

Fig. 2(e) and 2(f) show $L_B$ for $d_2$ = 712 nm (similar beatlengths are expected to $d_1$ = 36 μm exceptionally at resonances). TE$_{01}$ and TM$_{01}$ show similar $L_B$ deviating at shorter wavelengths (with TM$_{01}$ overlapping both $\Lambda$ curves). The acoustic period $\Lambda$, and its first harmonic, $\Lambda/2$, are calculated from $f$ = 40 to 130 kHz with (6) and compared to $L_B$. We note that for both cores, $L_B$ nearly matches $\Lambda/2$. Fig. 2(d) shows that the larger core causes reduced index difference, $n_{\mathrm{HE}} - n_{\mathrm{TE/TM}}$, improving the matching of $L_B$ and $\Lambda$ (Fig. 2(f)). This is suitable for increasing the modulation depth over a broad spectral range, as discussed further in Section IV.

## III. Experimental Setup and Characterization of the N-HCF Power Transmission Spectrum

Fig. 3 illustrates the AOM, and the experimental setup employed to characterize the N-HCF transmission spectrum. The AOM is composed of a piezoelectric transducer (PZT disc with 2 mm thickness and 25 mm in diameter), a solid silica horn (tapered from 1 mm to 270 μm along 3 cm), and a $L$ = 3.6 cm long N-HCF (with 5.1 cm total length including a coated segment used as an acoustic damper (green line in Fig. 3)). PZT, horn, and N-HCF ends are aligned and connected with a fixing adhesive by using XYZ micro stages and a microscope (M). The AOM is driven by an arbitrary signal generator (SG).

The N-HCF spectrum is first characterized without acoustic modulation from $\lambda$ = 750 to 1200 nm. This wavelength range has been chosen because the N-HCF exhibits low losses (Fig. 2(a)) and high-power transmission for core 2 (Fig. 4(d)) covering important wavelengths used in optical communications (980 nm), industrial and biomedical applications (1060 nm). In this case, we mention that even the smaller core 1 provides low losses and high transmission at these application wavelengths, as further shown in Fig. 4(c). In addition, the N-HCF has suitable optical and acoustic properties for acousto-optic modulation of core 2 in the considered spectral range, providing broad modulation bands, as further demonstrated in Fig. 5. The power of a supercontinuum source (SC) is aligned by mirrors and coupled to a single-mode fiber (SMF) by using an objective. The SMF is butt-coupled to the N-HCF, and the output power is collimated and further filtered with an iris (I) (adjusted to pass the power confined in each air core). The beam is further divided with a beamsplitter (BS): one beam passing through an attenuator (A) reaches a laser beam profiler (LBF), while the other beam is coupled to a multimode fiber (MMF) connecting to an optical spectral analyzer (OSA). The SMF core is carefully aligned to each N-HCF core, and its mode profile and spectrum are simultaneously measured.

Fig. 4(a) and 4(b) show the power distribution of the fundamental mode $HE_{11}$ in the N-HCF cores. As expected, the modes are mostly confined in the nearly circular region formed by the core silica edges (the orange and red curves show an almost Gaussian power distribution along the $y$-axis). This indicates that the unbent N-HCF favors the propagation of $HE_{11}$ rather than the higher-order modes, which could not be seen even by adjusting the coupling of SMF and N-HCF. Fig. 4(c) shows the measured and simulated spectra of $HE_{11}$ for the cores (considering only the nanoweb in the simulation). Fig. 4(c) evaluates the nanoweb's influence on the transmission properties of the N-HCF's cores. The simulations show that the nanoweb might induce high losses on the edges of the spectrum. However, the large difference in the distribution of the simulated and measured curves of core 2 indicates that the nanoweb does not significantly affect core 2 but causes high attenuation in core 1 at short wavelengths. The differences of the measured and simulated spectra of core 1 in Fig. 4(c) mostly evident at long wavelengths, might be caused by a power fraction of higher-order cladding modes, which could not be filtered with our iris (having a diameter larger than core 1). The filtering of this cladding power might reveal another high attenuation band from 1050 nm, as indicated by the simulation (dashed red). We note that both spectra of core 1 show a high magnitude around 950 nm (the resonances in the central band indicate the influence of the silica wall). Similarly, core 1 exhibits significantly high attenuation at shorter wavelengths, suggesting a stronger nanoweb's impact on the smaller core (nanoweb might also contribute to increased attenuation in core 2 at shorter $\lambda$). Considering that cores 1 and 2 have similar resonances in the central low attenuation range in Fig. 4(c), partial agreement of measured and simulated resonances is also expected for core 1. In contrast, Fig. 4(d) shows that core 2 provides well-defined resonances in the whole spectrum, caused by the larger overlap of the mode with the silica wall compared to that with the nanoweb, as shown in Fig. 4(b). The free spectral range (FSR) of measured and simulated resonances is shown in Fig. 4(d) (inset). Overall, the results indicate that a larger core might reduce the losses induced by the nanoweb broad resonances, at the expense of having narrow resonances in the entire range. Alternatively, the resonances in the N-HCF spectrum might be minimized by reducing the wall thickness $d$, providing a smoother transmission band (e.g., as indicated with the dashed blue curve in Fig. 4(c)).

Mode field distortion caused by the semi-circular core is ideally not expected [10]. However, field distortion deviating from a perfect circular Gaussian profile might be caused by the influence of higher-order modes. The off-center alignment between the laser and core might benefit a specific higher-order mode rather than just the fundamental mode, contributing to distortion. The results are obtained from the circular-core model, and the proposed dual-core N-HCF is mutually equivalent regarding the resonances (which depend on the fiber material and wall thickness in (1)) and the real part of the modes' effective index (considering the modes are mostly confined in the core edges). Overall, the semi-circular core might have slightly lower losses compared to a circular-core capillary. Differences in measured and simulated spectra might be caused by other loss contributions from absorption, scattering, and microbend [13]. In addition, the N-HCF ideal confinement losses might be lower than those estimated, due to the reduced mode-silica overlap in its non-circular cores compared to that in a capillary [10]. Minor changes in the actual N-HCF's material and cross-section dimensions along the fiber length might also contribute to deviations in acoustic and optical parameters. Overall, the agreement of measured and simulated results shows that the demonstrated simulations effectively predict and evaluate the modal and spectral properties of the N-HCF geometry with distinct cores and dimensions.

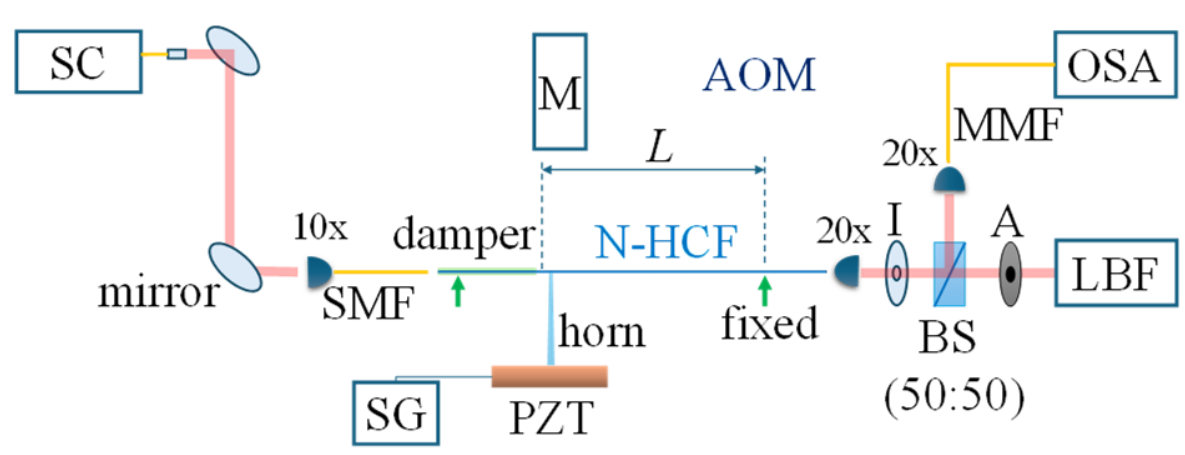

**Fig. 3.** Illustration of the experimental setup and the acousto-optic modulator (AOM) composed of a piezoelectric transducer (PZT), an acoustic horn, and an N-HCF of length $L$. The AOM is aligned with a microscope (M) and driven by a signal generator (SG). The modulated spectrum is characterized by using a supercontinuum source (SC), beamsplitter (BS), laser beam profiler (LBF), optical spectral analyzer (OSA), iris (I), attenuator (A), single mode and multimode fibers (SMF and MMF), and objectives.

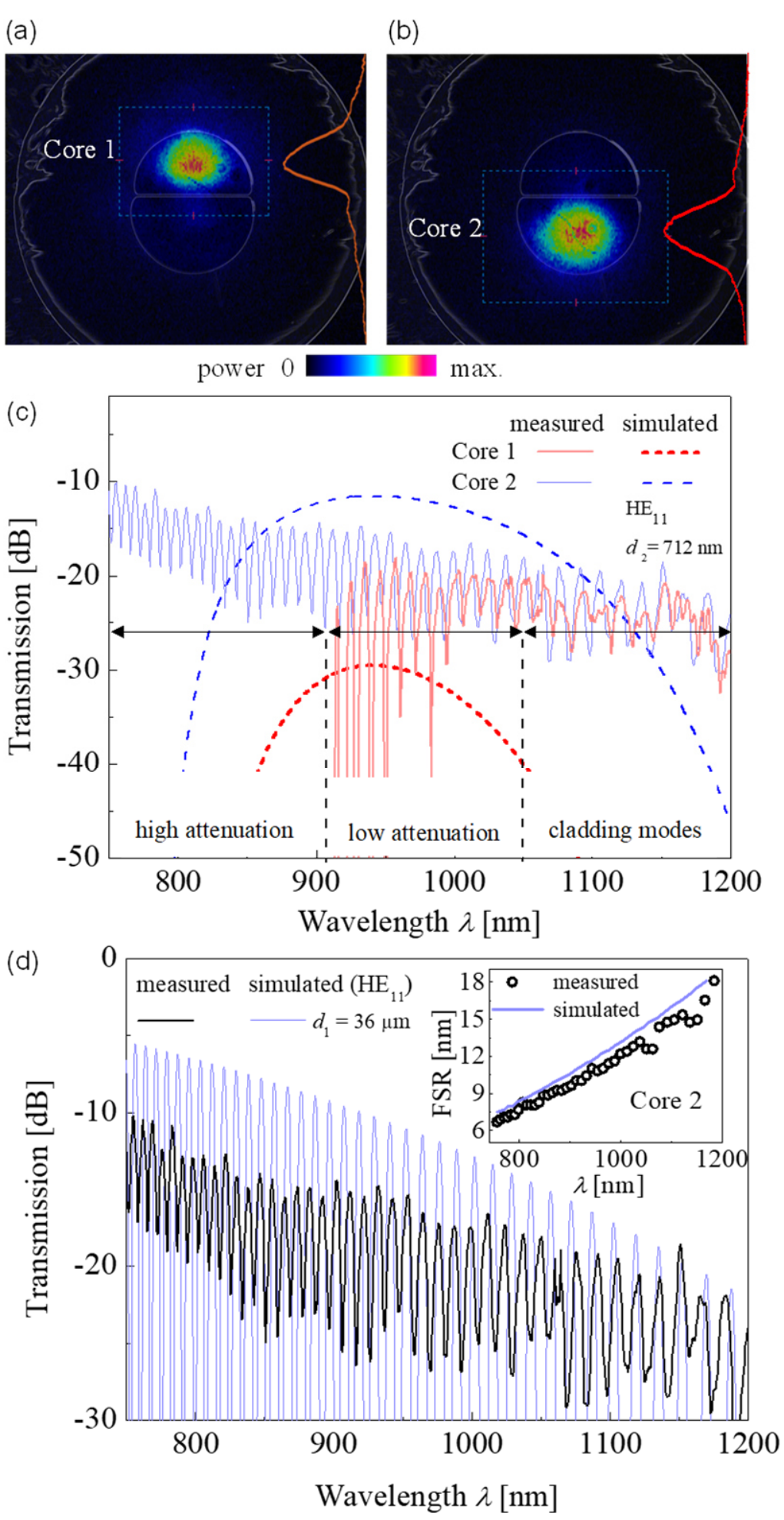

**Fig. 4.** Power distribution of the fundamental mode $HE_{11}$ in the (a) core 1 and (b) core 2 of the N-HCF (the orange and red curves on the right side show the normalized power distribution along the $y$-axis). (c) N-HCF's measured spectrum of core 1 and 2 (red and blue solid lines) compared to the simulated spectra (dashed lines) considering only the nanoweb effect ($d_2$ = 712 nm). (d) N-HCF's measured and simulated spectra of core 2 considering only the silica wall ($d_1$ = 36 μm). The inset shows the free spectral range (FSR) of the measured and simulated resonances.

## IV. Broadband Acousto-optic Modulation of the N-HCF Spectrum

We have characterized the modulated spectrum of the N-HCF by using the setup described in Section III (Fig. 3). PZT is tuned from $f$ = 40 to 100 kHz (1 kHz step) at a maximum of 10 V. The AOM works as an acoustic cavity inducing standing flexural acoustic waves at discrete resonant frequencies. Fig. 5 shows the N-HCF spectrum modulated by the strongest acoustic resonances. Fig. 5(a) shows the unmodulated (wave OFF) and modulated (wave ON) resonances of core 2 at $f$ = 71.5 kHz (the insets on the right show the mode's power distribution in the core for both cases). Fig. 5(b) shows the normalized modulated spectrum indicating only the effect of the acoustic wave. The power above the off-line might be caused by the resonances spectrally shifting with the acoustically changed refractive indices [16]. The 450 nm modulated bandwidth (3-dB) covers the wavelength range with significant modulation depth (up to 8 dB). This wide modulated band might be caused by the employed N-HCF short interaction length (3.6 cm) and its large beatlengths $L_B$ slightly changing in most spectrum in Fig. 2(f) [9]. In addition, other contributions might come from the simultaneous coupling of $HE_{11}$, $TE_{01}$, and $TM_{01}$. The higher modulation depths at shorter wavelengths denote strong coupling to $TM_{01}$ with $L_B$ approaching $\Lambda$ (other higher-order modes with similar effective indices might also be coupled). We have evaluated the AOM tunability by fine-tuning the acoustic frequency at $f$ = 71.1 kHz, as shown in Fig. 5(c). This tuning might highlight coupling for a specific mode (e.g., $TE_{01}$) (the profile of higher-order modes could not be seen due to their high attenuation). The notch envelope in Fig. 5(d) resembles the spectrum of a dynamically induced long-period grating (LPG), as those demonstrated in TL-HCFs [17]. Furthermore, the resonant wavelength $\lambda_C$ is tuned by adjusting $f$ over a maximum 80 nm range (Fig. 5(e)). Similarly, the modulation depth is tuned by the drive voltage, as shown in Fig. 5(f). Nevertheless, both nanoweb and off-center air cores jointly contribute to increasing the induced strain in the fiber cross-section, enhancing the modulation of core 2 [16]. Further studies might investigate N-HCFs with dissimilar cores for selective spectral filtering in AOMs or reduce the cross-sensitivity to measure multiple measurands in fiber sensors. For example, a smaller bend-insensitive core might be used to measure temperature, while a larger core might enhance sensitivity to mechanically induced deformations.

We have properly characterized core 1 but could not achieve any significant acousto-optic modulation because of the high losses and weak phase-matching indicated by the simulations (Fig. 2(e)). Therefore, we have analyzed the modulation only in core 2. This suggests that even a small core diameter difference of 5 μm is sufficient to affect both the N-HCF's optical and acoustic performance. The simulations show higher losses, larger differences in the effective indices of the coupling modes, and weaker phase-matching for core 1, which results in inefficient acousto-optic modulation. Thus, increasing core 1 might improve acousto-optic parameters as demonstrated for core 2. For two identical cores, the same performance demonstrated by core 2 is expected for core 1. For core 1 larger than core 2, the core might become highly multimode, and a new study is required to predict transmission and acousto-optic properties.

Although it is not demonstrated, the N-HCF might provide ideally lower losses compared to a conventional capillary with the same core diameter due to the reduced overlap between the mode field and the silica [10]. Considering the N-HCF and a capillary with the same outer diameter, the N-HCF core provides reduced silica content over the fiber cross section, benefiting from higher acousto-optic modulation efficiency [16]. This advantage increases with the increasing number of cores. Besides, N-HCF is significantly easier to fabricate compared to other anti-resonant hollow core fibers employing multiple concentric layers [11], kagomé structure [17] or tubular lattice (TL-HCF) [18], which might reduce costs and development time. The N-HCF has simpler geometry and a smaller outer diameter, facilitating connection and power coupling with other fiber components. Overall, N-HCF provides higher modulation efficiency (0.8 dB/V) and a broader modulation band compared to TL-HCFs [18], while employing a shorter fiber length. The acousto-optic modulation can change the fiber's dispersion characteristics because the acoustic wave changes the effective indices $n_{\text{eff}}$ of the modes [18]. Thus, chromatic dispersion can increase or even decrease depending on the altered $n_{\text{eff}}$ with the wavelength. Similarly, the bend-induced higher-order modes might also contribute to increasing the N-HCF modal dispersion. Overall, these induced dispersion effects might be negligible to practical applications, considering the quite short fiber used in the modulator (3.6 cm).

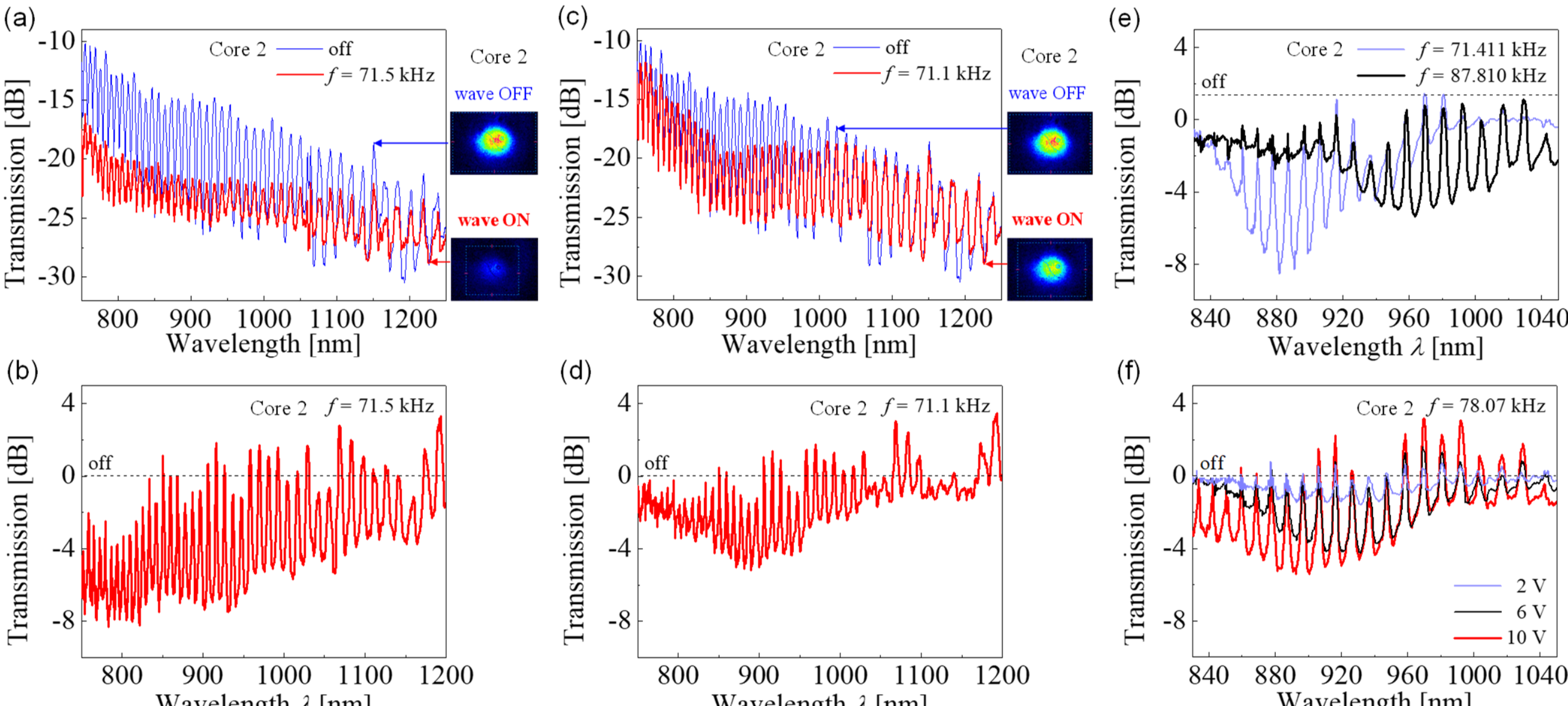

**Fig. 5**. N-HCF's measured unmodulated (wave OFF) and modulated (wave ON) transmission spectra at (a) $f$ = 71.5 kHz and (c) $f$ = 71.1 kHz (the spectra are respectively normalized in (b) and (d)). Tuning the (e) center wavelength and (f) depth of modulated notches by changing the drive frequency and voltage.

In summary, this study provides the following contributions: (a) First experimental demonstration of an acoustically modulated dual-core N-HCF; (b) Widely tunable modulated bandwidths (80 – 450 nm) covering from 750 to 1200 nm, which is useful to shorten the pulse width of pulsed fiber lasers; (c) High modulation efficiencies (up to 0.8 dB/V) with no use of tapering or etching, which is comparable with those using etched fibers and TL-HCFs, and higher than those with SMFs [17]; (d) Compact 3.6 cm long device promising to reduce the response time and size of AOMs (~54 % smaller [17]); (e) The demonstrated analytical simulations effectively evaluate the N-HCF optical and acoustic parameters, which is promising for the design, modeling, and analysis of HCFs with distinct geometries, core diameters, and wall thicknesses.

## V. Conclusion

We demonstrate the acousto-optic modulation of a dual-core antiresonant N-HCF for the first time. To the best of our knowledge, the 450 nm modulated bandwidth covering from 750 to 1200 nm is the broadest ever achieved. N-HCF is analytically simulated, and the influence of the cores' diameters ($D$ = 19 and 23 μm), silica wall, and nanoweb thicknesses ($d_1$ = 36 μm and $d_2$ = 712 nm) on modal and acoustic properties are evaluated. The measured and simulated results indicate that the distinct thickness of the wall and nanoweb is negligible for the confinement losses around 980 nm. However, nanoweb might work as a passband filter in core 1, inducing notable attenuation at shorter wavelengths. The smaller core diameter provides higher losses and irrelevant acoustic modulation. In contrast, core 2 shows lower losses and higher dependency on the wall by enhancing narrow resonances in the spectrum. Overall, both N-HCF cores and nanoweb strongly contribute to increased modulation strength, improving the AOM's efficiency. The fabricated 3.6 cm long device offers significantly high modulation depths (up to 8 dB) at a low drive voltage (10 V). These features are promising for compact, fast, and efficient fiber sensors, spectral filters, and modulators for ultrashort pulsed fibers.

## Acknowledgment

The authors thank Claudenete Vieira Leal with LAMULT for assisting with the microscope images of the N-HCF. We thank the LIMicro-IQ – Microscopy Core Facility (RRID:SCR_024633) at UNICAMP for support. The Quanta FEG 250 system was partially funded by a FAPESP grant (#2023/01620-4) for the LIMicro-IQ Core Facility.